\documentclass{iaus}
\usepackage{graphicx}

\newcommand{\EQ}{\begin{equation}}
\newcommand{\EN}{\end{equation}}
\newcommand{\EQA}{\begin{eqnarray}}
\newcommand{\ENA}{\end{eqnarray}}

\newcommand{\EEq}[1]{Equation~(\ref{#1})}

\newcommand{\Fig}[1]{Figure~\ref{#1}}

\newcommand{\bra}[1]{\langle #1\rangle}

{}
{}
{}

{}
{}
{}
{}
{}
{}
{}
{}
{}
{}
{}
{}
{}
{}
{}
{}
{}
{}

{}
{}
{}

{}

{}
{}

\newcommand{\UU}{\mbox{\boldmath $U$} {}}

\newcommand{\BB}{\mbox{\boldmath $B$} {}}

\newcommand{\JJ}{\mbox{\boldmath $J$} {}}

\newcommand{\FF}{\mbox{\boldmath $F$} {}}

\newcommand{\nab}{\mbox{\boldmath $\nabla$} {}}
\newcommand{\DD}{{\rm D} {}}

\def\kf{k_{\rm f}}

\newcommand{\yana}[3]{ #1, {A\&A,} {#2}, #3}

\newcommand{\ybook}[3]{ #1, {#2} (#3)}

\title[Recurrent flux emergence from dynamo-generated fields] %% give here short title %%
{Recurrent flux emergence from dynamo-generated fields}

\author[J{\"o}rn Warnecke \& Axel Brandenburg]   %% give here short author list %%
{J{\"o}rn Warnecke$^{1,2}$
 \and Axel Brandenburg$^{1,2}$}

\affiliation{$^1$Nordita, AlbaNova University Center, \\Roslagstullsbacken 23,
SE-10691 Stockholm, Sweden \\email: {\tt joern@nordita.org} \\[\affilskip]
$^2$Department of Astronomy, AlbaNova University Center, \\Stockholm University, 
SE 10691 Stockholm, Sweden}

\pubyear{2010}
\volume{271}  %% insert here IAU Symposium No.
\pagerange{119--126}
\jname{Astrophysical Dynamics: From Stars to Galaxies}
\editors{N. Brummell \& A.S. Brun}
\begin{document}

\maketitle

\begin{abstract}
We investigate the emergence of a large-scale magnetic field. This field is 
dynamo-generated by turbulence driven with a helical forcing function.
Twisted arcade-like field structures 
are found to emerge in the exterior above the turbulence zone.
Time series of the magnetic field structure show recurrent plasmoid 
ejections.

\end{abstract}

\firstsection % if your document starts with a section,
\section{Introduction}

The magnetic field at the visible surface of the Sun is known to take
the form of bipolar regions and the field continues in an arch-like fashion.
These formations appear usually as twisted loop-like structures.
These loops can be thought of as a continuation of more concentrated
flux ropes in the bulk of the solar convection zone.
Twisted magnetic fields are produced by a large-scale dynamo mechanism
that is generally believed to be the motor of solar activity (Parker 1979).
One such dynamo mechanism is the $\alpha$ effect that produces a large-scale
poloidal magnetic field from a toroidal one.
In order to study the emergence of helical magnetic fields from a
dynamo, we consider a model that combines a direct simulation of a
turbulent large-scale dynamo with a simple treatment of the evolution
of nearly force-free magnetic fields above the surface of the dynamo.
In the context of force-free magnetic field extrapolations this method is 
also known as the stress-and-relax method (Valori et al.\ 2005).
Above the solar surface, we expect the magnetic fields to drive flares and
coronal mass ejections through the Lorentz force.
In the present paper we highlight some of the main results of our earlier
work (Warnecke \& Brandenburg 2010).

\section{The Model}
The equation for the velocity correction in the Force-Free Model is
similar to the usual momentum equation,
except that there is no pressure, gravity, or other driving forces on
the right-hand side, so we just have
\begin{equation}
{\DD\UU\over\DD t}=\JJ\times\BB/\rho+\FF_{\rm visc},
\label{DUDtext}
\end{equation}
where $\JJ\times\BB$ is the Lorentz force,
$\JJ=\nab\times\BB/\mu_0$ is the current density,
$\mu_0$ is the vacuum permeability,
$\FF_{\rm visc}$ is the viscous force,
and $\rho$ is here treated as a constant
the determines the strength of the velocity correction.
\EEq{DUDtext} is solved together with the induction equation.
In the lower layer the velocity is excited by a forcing function and the density
is evolving using the continuity equation.
The forcing function consists of random 
plane helical transversal waves with an average forcing wavenumber $\kf$.
\section{Results}
\begin{figure}[t!]\begin{center}
\includegraphics[width=7cm]{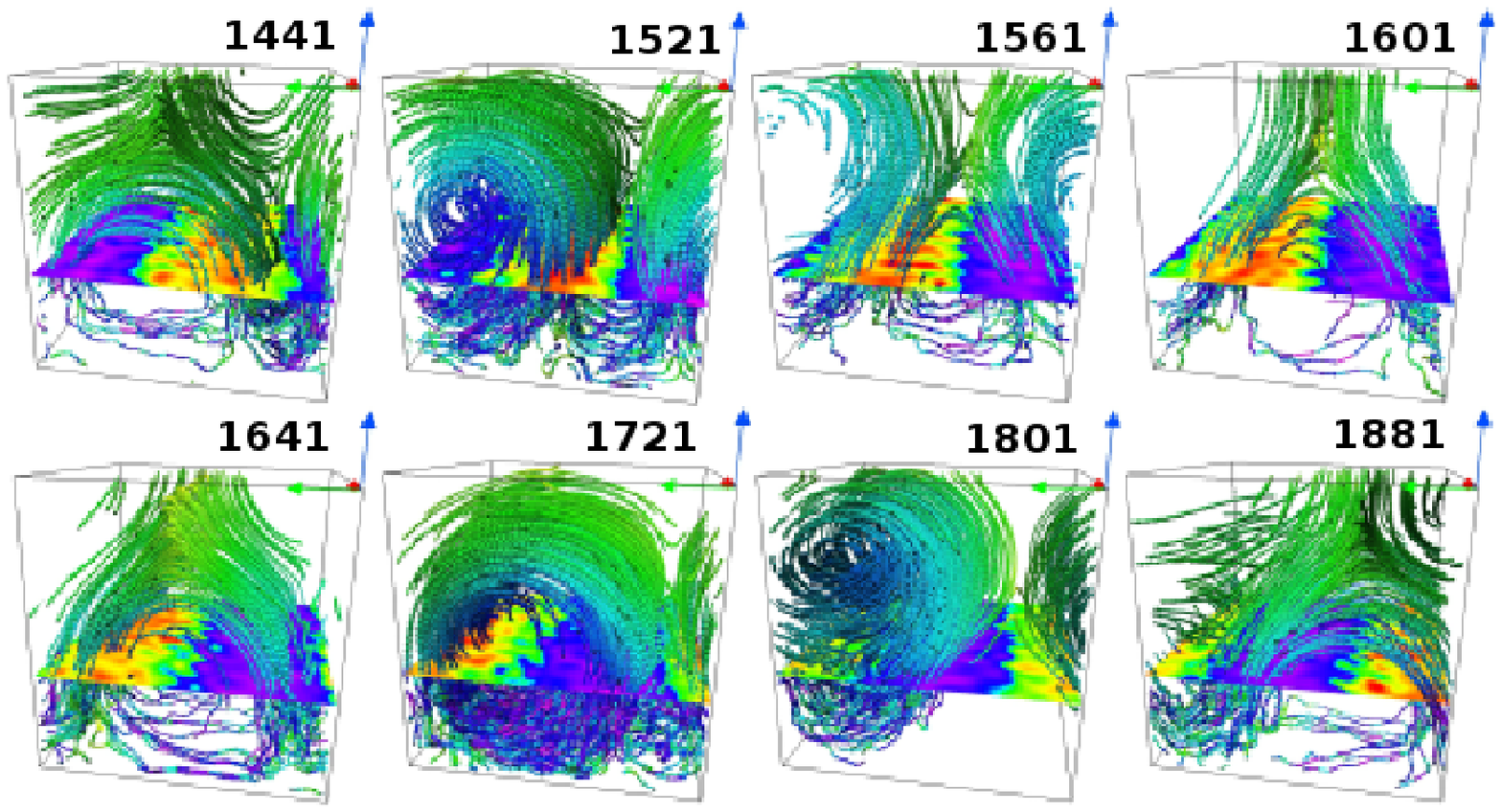}
\includegraphics[width=6.3cm]{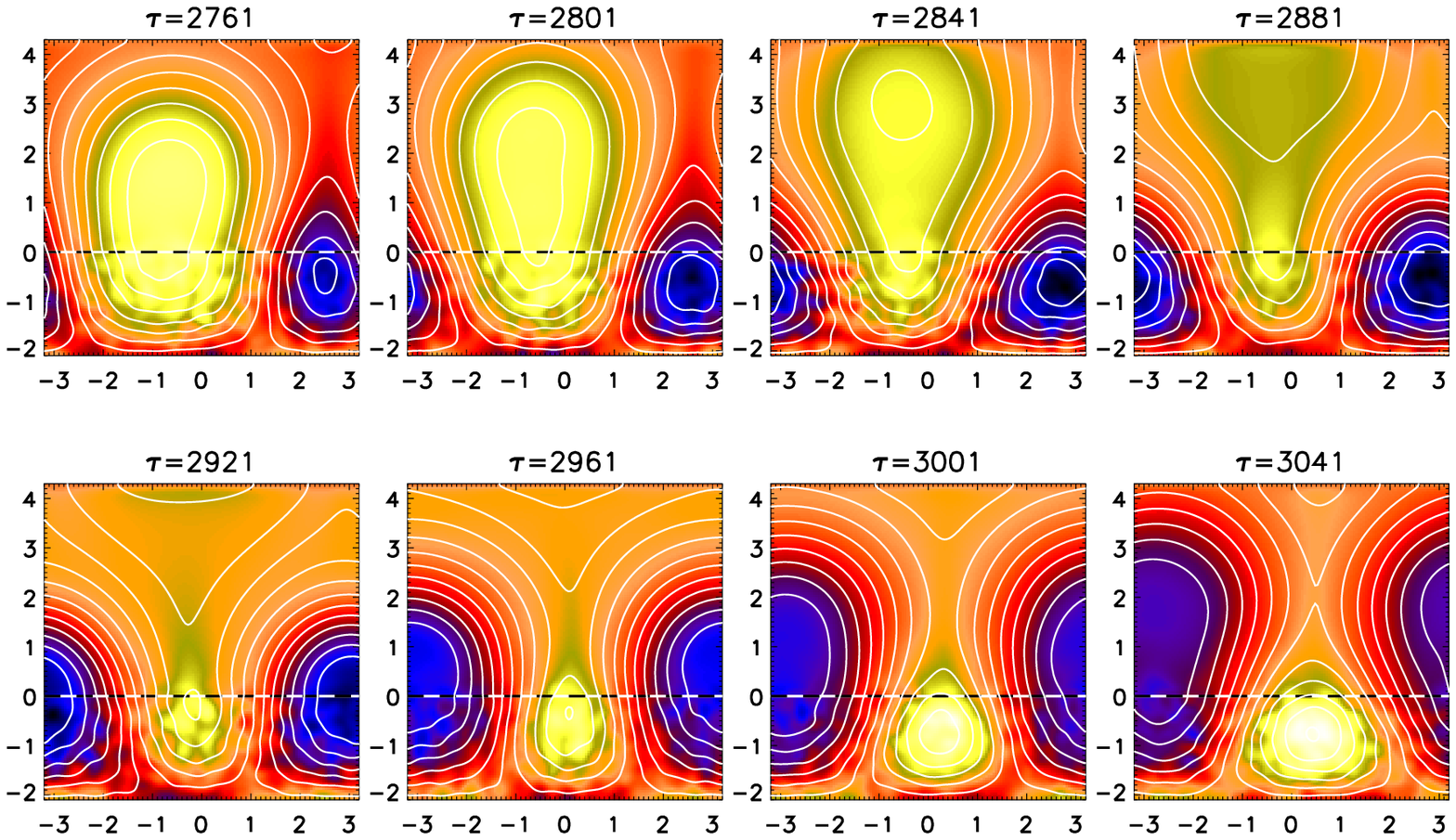}
\end{center}\caption[]{
{\it Left panel}: Time series of arcade formation and decay. Field lines are colored by 
their local field strength which increases from pink to green.
The plane shows $B_z$ increasing from red (positive) to pink (negative).
The normalized time $\tau$ is giving in each panel.
{\it Right panel}: Time series of the formation of a plasmoid ejection. Contours of 
$\bra{A_x}_{x}$ are shown together with a color-scale representation of $\bra{B_x}_x$;
dark blue stands for negative and red for positive values.
The contours of $\bra{A_x}_{x}$ correspond to field lines of $\bra{\BB}_x$
in the $yz$ plane.The dotted horizontal lines show the location of the surface at $z=0$.
Adapted from Warnecke \& Brandenburg (2010).
}
\label{AI}
\end{figure}

The magnetic field grows first exponentially and then shows subsequent
saturation that is typical for forced turbulent dynamo action.
In the turbulent layer the magnetic 
field reaches around $78\%$  of the equipartition field strength, $B_{\rm eq}$.
The dynamo generates
a large-scale field whose vertical component has a sinusoidal 
variation in $y$. After some time the magnetic field extends well into the exterior 
where it tends to produce an arcade-like structure, as seen in the left panel
of \Fig{AI}. The arcade opens up in the middle above the line where the vertical field 
component vanishes at the surface. This leads to the formation of anti-aligned field 
lines with a current sheet in the middle.
The dynamical evolution is seen clearly in a sequence of field line images
in the left hand panel 
of \Fig{AI}, where anti-aligned vertical field lines reconnect above the neutral line 
and form a closed arch with plasmoid ejection above. This arch then changes its 
connectivity at the foot points in the sideways direction (here the $y$ direction), 
making the field lines bulge upward to produce a new reconnection site with anti-aligned 
field lines some distance above the surface.
Field line reconnection is best seen for two-dimensional magnetic fields, because it is 
then possible to compute a flux function whose contours correspond to field lines in the 
corresponding plane.
In the present case the large-scale component of the
magnetic field varies only little in the
$x$ direction, so it makes sense to visualize the field averaged in the $x$ direction.
The right panel of \Fig{AI} shows clearly the recurrent reconnection events with
subsequent plasmoid ejection.
The dynamics of the magnetic
field in the exterior is indeed found to mimic open boundary conditions
at the interface between the turbulence zone and the exterior at $z=0$.
In particular, it turns out that a twisted magnetic field generated
by a helical dynamo beneath the surface is able to produce flux emergence
in ways that are reminiscent of that found in the Sun.

\end{document}